\documentclass[letterpaper, 10pt, conference]{ieeeconf}

\IEEEoverridecommandlockouts
\newcommand\beqa{\begin{eqnarray}}
\newcommand\eeqa{\end{eqnarray}}

\usepackage{cite}
\usepackage{amsmath,amssymb,amsfonts}

\usepackage{amsthm}

\let\labelindent\relax
\usepackage{algorithmic,enumitem}

\usepackage{graphicx}
\usepackage{subcaption}
\usepackage{textcomp}
\usepackage{xcolor}
\usepackage{booktabs}

\newtheorem{theorem}{Theorem}[section]
\newtheorem{lemma}[theorem]{Lemma}

\theoremstyle{remark}

\title{\LARGE \bf
Distributed Attraction-Repulsion Potential for \\
Multi-Agent Formation Control
}

\author{Hemanta Ban$^{1}$, Seddik M. Djouadi$^{1}$, Kevin Tomsovic$^{2}$%
\thanks{$^{1}$ Hemanta Ban and Seddik M. Djouadi are with the Department of
        Electrical Engineering and Computer Science,
        University of Tennessee, Knoxville, TN 37996-2250, USA.
        {\tt\small hban@vols.utk.edu, mdjouadi@utk.edu}}%
\thanks{$^{2}$ Kevin Tomsovic is with Clemson University, Charleston Innovation
        Campus, 1240 Supply Street, North Charleston, SC 29405, USA.
        {\tt\small ktomsov@clemson.edu}}%
}

\begin{document}

\maketitle
\thispagestyle{empty}
\pagestyle{empty}

\begin{abstract}

In this paper, a distributed multi-agent formation control driven by the gradient
of the Lennard-Jones potential is analyzed. For a collision-free initial data
we prove global well-posedness together with a uniform lower bound on all
inter-agent distances, thereby excluding hard collisions. Taking the total
energy as a Lyapunov function, LaSalle's invariance principle shows that every
positive limit point is an equilibrium. Since trajectories remain uniformly
away from collisions, the energy is analytic along the flow and an argument
yields convergence to a single equilibrium modulo translations. Illustrative
numerical examples are presented.

\end{abstract}

\section{INTRODUCTION}

The coordination and control of multi-agent systems (MAS) have become a central
focus in modern control theory due to their broad applicability in areas such as
remote sensing, distributed computation, autonomous vehicle coordination, and
complex infrastructure monitoring \cite{OlfatiSaber2004, Nagahara2024}.
Coordinated operation of multiple autonomous agents offers advantages over
single-agent systems in robustness, capability, and cost efficiency. A key
challenge is to design distributed control laws that achieve coordination using
only local sensing, communication, and computation among neighboring agents,
without centralized supervision \cite{Jadbabaie2003}.

Within this framework, formation control is a fundamental coordination problem
that aims to drive multiple agents to satisfy prescribed state constraints and
achieve a desired spatial configuration \cite{Ahn2020}. Formation control
approaches are commonly classified as position-based (requiring global
coordinates), displacement-based (requiring aligned local frames), or
distance-based (using only relative distances in misaligned frames)
\cite{Oh2015, Liu2023}. We focus on the distance-based paradigm, which
minimizes reliance on global information but introduces significant nonlinear
dynamics challenges.

The control potential investigated in this paper is a distributed
attraction-repulsion model derived from physically inspired interaction laws
used in computational geometry and molecular dynamics
\cite{Zheleznyakova2011, Shimada1993}. The pair interaction potential
$\phi(\cdot)$ employed in our formulation is the Lennard-Jones (LJ) 12-6
potential \cite{ZhangSmirnov2005}, which provides short-range dominant
repulsion and longer-range weaker attraction \cite{Nie2010}. This structure
governs the distributed control input of each agent through the relative
distance $r_{ij}$ and the desired distance $\sigma_{ij}$.

The use of attraction-repulsion potentials is well motivated in the literature.
In particular, \cite{GaziPassino2004} introduced a general class of such
functions for stable swarm aggregation and formation control, while
\cite{OlfatiSaber06} developed a flocking framework based on gradient-type
interaction rules and velocity consensus.

To rigorously analyze such potential-based MAS, we rely on nonlinear stability
tools, particularly Lyapunov functions and LaSalle's invariance principle
\cite{Ogren2001, LaSalle1960}, to characterize asymptotic behavior. However,
deploying robust potential-based formation control laws faces two persistent
analytical challenges:

\begin{itemize}
    \item[] i. Collision Avoidance and Global Well-Posedness
\end{itemize}

In distributed systems, collision avoidance is paramount. Although repulsive
interaction laws can discourage collisions, they do not by themselves guarantee
global well-posedness or exclusion of hard collisions throughout the flow
\cite{OlfatiSaber06}. The LJ potential is attractive in this regard because its
repulsive force becomes unbounded as $r_{ij} \to 0$, but rigorous proofs that
collision-free initial conditions remain uniformly away from collisions for all
time are often absent in gradient-like settings.

\begin{itemize}
    \item[] ii. Convergence to a Unique Equilibrium
\end{itemize}

For gradient-based control laws, convergence is typically established only to a
set of equilibria ($\Omega_e$). Demonstrating convergence to a unique desired
equilibrium state, modulo trivial motions such as translations or rotations,
remains challenging for general $n$-agent formations because arbitrary
interaction topologies may admit non-trivial undesired equilibria.

This paper addresses both challenges through a rigorous analysis of a
distributed MAS driven by an LJ-based attraction-repulsion potential. We prove
global well-posedness and establish a uniform positive lower bound on all
inter-agent distances for every collision-free initial condition, providing a
non-local collision-exclusion guarantee. We then show that the total energy is
a Lyapunov function and, via LaSalle's invariance principle, that every
positive-limit point is an equilibrium. Because the trajectories remain
uniformly away from collisions, the energy is analytic along the flow, which
enables arguments based on the Lojasiewicz inequality \cite{Lojasiewicz1965}
and yields convergence to a single equilibrium modulo translations. This
resolves the ambiguity often associated with convergence in gradient-like
systems.

These analytical results provide a global stability and safety characterization
for this class of distributed, potential-driven MAS. The effectiveness of the
approach is illustrated through a numerical example.

The paper proceeds as follows: Section II introduces the system model and
establishes basic properties. Section III presents the Lyapunov-based
convergence analysis and characterizes the equilibrium set. Section IV
demonstrates the practical effectiveness of the proposed approach through
implementation details and numerical experiments. Finally, Section V concludes
the paper and discusses future directions.

\section{MODEL AND BASIC PROPERTIES}

In the physics-based node placement methods proposed in
\cite{ZhangSmirnov2005, Nie2010, Shimada1993, Zheleznyakova2011},
the distribution of the nodes is governed by the following equations of motion
for node $i$ given as
\begin{equation}
m_i\,\ddot{\mathbf{x}}_i(t) + c_i\,\dot{\mathbf{x}}_i(t) = \mathbf{F}_i(t),
\qquad i=1,\dots,N
\label{eq1}
\end{equation}
where
\begin{itemize}
  \item $m_i$ denotes the mass of agent $i$.
  \item $c_i$ is the damping coefficient of agent $i$, i.e.,
        $c_i\,\dot{\mathbf{x}}_i(t)$ is the damping term of node $i$.
  \item $\mathbf{x}_i(t)$ is the position of node $i$, the center of agent
        $i$, at time $t$, with $\mathbf{x}_i(t)\in \mathbb{R}^d$, $d=2$ or $3$.
  \item $\mathbf{F}_i(t)$ is the sum of the net forces acting on node $i$.
\end{itemize}
\begin{equation}
\mathbf{F}_i(t) = -\sum_{j\neq i} f(w_{ij})\,\mathbf{r}_{ij}(t)
\label{eq2}
\end{equation}
where $w_{ij} := \frac{r_{ij}}{\sigma_{ij}}$,
$r_{ij} := \|\mathbf{x}_i - \mathbf{x}_j\|$,
$\sigma_{ij} := \frac{q_i+q_j}{2}$.

The function $r_{ij}$ is the distance between node $i$ and node $j$, i.e., the
actual distance between the centers of agents $i$ and $j$, while $q_i$ and
$q_j$ are the radii of bubbles (agents) $i$ and $j$, respectively. The
distance $\sigma_{ij}$ is the distance at which the two agents $i$ and $j$ are
just touching without overlap \cite{Shimada1993}, and
\begin{equation}
\mathbf{r}_{ij}(t) :=
\frac{\mathbf{x}_i(t) - \mathbf{x}_j(t)}{\|\mathbf{x}_i(t) - \mathbf{x}_j(t)\|}
\label{eq6}
\end{equation}

In \cite{ZhangSmirnov2005}, it is proposed to solve the bubble meshing node
placement problem by using the potential energy function
\begin{equation}
U(\mathbf{x}_1, \mathbf{x}_2, \ldots, \mathbf{x}_N) :=
\sum_{i=1}^{N-1} \sum_{j=i+1}^{N}
\phi\!\left(\sigma_{ij}, \|\mathbf{x}_i - \mathbf{x}_j\|\right)
\label{eq7}
\end{equation}
The function $\phi(\cdot)$ corresponds to the LJ-potential between nodes $i$
and $j$ given by
\begin{equation}
\phi(\sigma_{ij}, r_{ij}) := 4a \left[
\left(\frac{\sigma_{ij}}{r_{ij}}\right)^{12} -
\left(\frac{\sigma_{ij}}{r_{ij}}\right)^{6}
\right]
\label{eq8}
\end{equation}
where $a$ is a positive constant representing the depth of the potential well,
i.e., it controls the strength of attraction of the agents. The term
$(\sigma_{ij}/r_{ij})^{12}$ represents the short-range repulsion, while
$(\sigma_{ij}/r_{ij})^{6}$ represents the attractive part.
$\phi(\sigma_{ij}, r_{ij})$ as a function of $r_{ij}$ is continuously
differentiable on $(0,\infty)$, with
$\phi(\sigma_{ij}, r_{ij}) \to +\infty$ as $r_{ij} \downarrow 0$,
and $\phi(\sigma_{ij}, r_{ij}) \to 0^-$ as $r_{ij} \to \infty$.

The potential function $U(\cdot)$ is chosen such that it is zero whenever the
agents are at the desired position and positive otherwise. In
\cite{ZhangSmirnov2005}, the agents' motion obeys the steepest descent control
law along the gradient $\nabla U(\cdot)$, more precisely
\begin{equation}
f_{ij} := f(w_{ij}) = -\frac{d\phi(\sigma_{ij}, r_{ij})}{dr_{ij}}
= 4a\left[\frac{12\sigma_{ij}^{12}}{r_{ij}^{13}} -
          \frac{6\sigma_{ij}^{12}}{r_{ij}^{7}}\right]
\label{eq9}
\end{equation}
Note that $f_{ij}=f_{ji}$ and the equilibrium, i.e., $f_{ij}\equiv 0$, is
reached at \cite{Rapaport2004},
\begin{equation}
\sigma_{0,ij} = 2^{\tfrac{1}{6}}\sigma_{ij}
\label{eq10}
\end{equation}
As $r_{ij}$ increases towards $2^{\tfrac{1}{6}}\sigma_{ij}$, $f_{ij}$
decreases to zero. $2^{\tfrac{1}{6}}\sigma_{ij}$ is the unique minimizer for
$\phi(\sigma_{ij}, r_{ij})$, since
\begin{equation}
\frac{d^2\phi}{dr_{ij}^2}\!\left(\sigma_{ij}, 2^{\tfrac{1}{6}}\sigma_{ij}\right)
= \frac{72a}{2^{\tfrac{1}{3}}\sigma_{ij}^2} > 0
\label{eq11}
\end{equation}
The minimum is then
\begin{equation}
\phi\!\left(\sigma_{ij}, 2^{\tfrac{1}{6}}\sigma_{ij}\right) = -a
\label{eq12}
\end{equation}
It is important to notice that the net force $\mathbf{F}_i(t)$ is as expected
given by
\begin{equation}
\mathbf{F}_i(t) = -\nabla_{\mathbf{r}_{ij}}
U(\mathbf{x}_1, \mathbf{x}_2, \cdots, \mathbf{x}_n)
\label{eq13}
\end{equation}
Let us write equations (\ref{eq1}) and (\ref{eq2}) in state space form by
first setting for agent $i$,
\begin{equation}
\mathbf{z}_i =
\begin{bmatrix} \mathbf{x}_i \\ \dot{\mathbf{x}}_i \end{bmatrix}
\in \mathbb{R}^{2d}
\label{eq14}
\end{equation}
The second-order dynamics (\ref{eq1}) become the first-order nonlinear state
space system for $i=1, 2, \cdots, N$,
\begin{align}
\dot{\mathbf{z}}_i
=
\begin{bmatrix}
\dot{\mathbf{x}}_i \\[6pt]
\ddot{\mathbf{x}}_i
\end{bmatrix}
=
\begin{bmatrix}
\mathbf{v}_i \\[8pt]
\dfrac{1}{m_i}\!\left(-c_i\,\mathbf{v}_i
- \displaystyle\sum_{j \neq i} f_{ij}(w_{ij})\,\mathbf{r}_{ij}\right)
\end{bmatrix}
\label{eq15}
\end{align}
with initial state for $i=1,2,\cdots,N$,
\begin{equation}
\mathbf{z}_i(0) =
\begin{bmatrix} \mathbf{x}_i(0) \\ \dot{\mathbf{x}}_i(0) \end{bmatrix}
\in \mathbb{R}^{2d}
\label{eq12a}
\end{equation}

We assume that $\mathbf{z}_i(0)$ satisfies $r_{ij}(0) > 0$ for all $i \neq j$,
so the initial potential
$U(\mathbf{x}_1(0), \mathbf{x}_2(0), \dots, \mathbf{x}_N(0))$ is
well-defined.

The equilibrium points of this system correspond to
$\dot{\mathbf{x}}_i = \mathbf{v}_i = 0$,
\begin{equation}
\nabla_{\mathbf{x}_i} U =
\sum_{j \neq i} f_{ij}(w_{ij})\,\mathbf{r}_{ij} = 0,
\quad i = 1,2,\dots,N
\label{eq17}
\end{equation}
Equivalently,
\begin{equation}
\sum_{j\neq i} 24a \left[
2\left(\frac{\sigma_{ij}}{r_{ij}}\right)^{12}
- \left(\frac{\sigma_{ij}}{r_{ij}}\right)^{6}
\right] \frac{\mathbf{r}_{ij}}{r_{ij}} = 0
\label{eq18a}
\end{equation}

Call the set of equilibria
$\Omega_e := \{(\mathbf{x}, \mathbf{v}): \mathbf{v}=0,\;
\nabla_{\mathbf{x}} U = 0\}$,
where
$\mathbf{x} := (\mathbf{x}_1, \cdots, \mathbf{x}_N)^T \in \mathbb{R}^{dN}$
and
$\mathbf{v} := (\mathbf{v}_1, \cdots, \mathbf{v}_N)^T \in \mathbb{R}^{dN}$.
If $(\mathbf{x}_e, 0)$ is an equilibrium point, then its translate
$(\mathbf{x}_e + \mathbf{c}, 0)$ is also an equilibrium point for any constant
vector $\mathbf{c} \in \mathbb{R}^{dN}$, since $U(\cdot)$ is only a function of
the distances $r_{ij}$. The statement that $(\mathbf{x}(t), \mathbf{v}(t))$
converges to $(\mathbf{x}_e, 0) \in \Omega_e$ as $t \to \infty$ means
convergence modulo translations.

Note for two agents $N=2$, from (\ref{eq10}) the equilibria occur at
$f_{12}(w_{12})=0$, i.e., $r_{12} = 2^{\tfrac{1}{6}}\sigma_{12}$. For more
agents $N > 2$, that is not the case anymore and the vectors involved in
(\ref{eq17}) must vectorially add to zero. However, the agent motions converge
to the minimum potential configuration.

We consider three particles with positions with $\sigma_{ij} = \sigma$,
\[
x_1 = (0,0), \quad x_2 = (L,0), \quad
x_3 = \left(\tfrac{L}{2},\tfrac{\sqrt{3}}{2}L\right).
\]
The distances are $r_{12} = r_{23} = r_{31} = L$.
The force magnitude vanishes when
\begin{equation}
\frac{d\phi(\sigma,r)}{dr}
= 24a\left[-2\frac{\sigma^{12}}{r^{13}} + \frac{\sigma^6}{r^7}\right] = 0
\label{eq:N}
\end{equation}
which occurs at $r = 2^{1/6}\sigma$. Setting $L = 2^{1/6}\sigma$ makes all
three distances equal the zero-force distance. By symmetry of the equilateral
triangle the vector sum of forces on each bubble cancels, so
$\nabla_{\mathbf{x}} U = 0$ and the equilateral triangle of side
$2^{1/6}\sigma$ is an equilibrium configuration.

Again for $N=3$, but with agents' centers collinear,
\[
x_1 = (-b,0), \quad x_2 = (0,0), \quad x_3 = (b,0).
\]
Thus $r_{12} = r_{23} = b$ and $r_{13} = 2b$. It can be shown (details
omitted) that the equilibrium configuration is
\[
x_1 = (-1.1213\sigma,0), \quad x_2 = (0,0), \quad x_3 = (1.1213\sigma,0).
\]
Note that
$r_{12} = r_{23} \approx 1.1213\sigma$ and $r_{13} \approx 2.2426\sigma$.
Neither $r_{12}$ nor $r_{13}$ equals the $N=2$ equilibrium distance
$\sigma_{0,ij} = 2^{1/6}\sigma$. Instead, the repulsion from the middle agent
balances the attraction from the far outer agent, showing that in an $N=3$
system equilibrium need not correspond to placing all pairs at $\sigma_{0,ij}$.

The total energy $E(\cdot)$ of the system (\ref{eq15}) consists of kinetic
energy $E_K(\cdot)$ and potential energy $U(\cdot)$,
\begin{equation}
E(\mathbf{x}(t)) = E_K(\mathbf{x}(t)) + U(\mathbf{x}(t))
\label{eq16}
\end{equation}
\begin{equation}
E_K(\mathbf{x}(t)) :=
\frac{1}{2}\sum_{i=1}^N m_i\|\dot{\mathbf{x}}_i(t)\|^2
\label{eq_EK}
\end{equation}
with initial total energy
\begin{equation}
E_0 := E(\mathbf{x}(0)) =
\frac{1}{2}\sum_{i=1}^N m_i\|\dot{\mathbf{x}}_i(0)\|^2 + U(\mathbf{x}(0))
\label{eq18}
\end{equation}
Define the collision set
\begin{equation}
\mathcal{C} :=
\{\mathbf{x} \in \mathbb{R}^{dN} : r_{ij} = 0,\; i \neq j\}
\label{eq19}
\end{equation}
The total energy $E$ is nonincreasing along trajectories since
\begin{align}
\frac{dE}{dt} = \frac{d(E_K + U)}{dt}
&= \sum_{i=1}^N \left(m_i\ddot{\mathbf{x}}_i^T
   + \nabla_{\mathbf{x}_i} U^T\right)\dot{\mathbf{x}}_i \nonumber \\
&= -\gamma\sum_{i=1}^N \|\dot{\mathbf{x}}_i\|^2 \leq 0
\label{eq20}
\end{align}
This shows that $E \leq E_0$ for all $t \geq 0$, keeping $E$ finite. From
(\ref{eq8}) and (\ref{eq9}) it can be shown (details omitted due to space
limitations) that with $\sigma_{\min} := \min_{i < j}\sigma_{ij}$,
\begin{equation}
U \geq 2a\sigma_{\min}^{12}
\sum_{i=1}^{N-1}\sum_{j=i+1}^{N} \frac{1}{r_{ij}^{12}}
- 2a\binom{N}{2}
\label{eq21}
\end{equation}
Since $E_0 \geq E \geq U$, it follows
\begin{equation}
\frac{1}{r_{ij}^{12}} \leq
\sum_{i=1}^{N-1}\sum_{j=i+1}^{N}\frac{1}{r_{ij}^{12}}
\leq \frac{E_0 + 2a\binom{N}{2}}{2a\sigma_{\min}^{12}} > 0
\label{eq22}
\end{equation}
Hence,
\begin{equation}
r_{\min}(t) := \min_{i<j} r_{ij}(t) \geq
\left(\frac{2a\sigma_{\min}^{12}}{E_0 + 2a\binom{N}{2}}\right)^{1/12} > 0
\label{eq23}
\end{equation}
This shows that $r_{ij}(t)$ is always bounded below by the RHS of (\ref{eq23})
and never reaches $0$. It follows that the vector field in (\ref{eq15}) is in
$C^\infty$ on the open set $\mathbb{R}^d / \mathcal{C}$, the complement of
$\mathcal{C}$. The bound in (\ref{eq23}) guarantees that the trajectories of
(\ref{eq2}) never exit $\mathbb{R}^d / \mathcal{C}$, and therefore the
solutions to (\ref{eq15}) for $i = 1,2,\dots,N$ exist and are unique, defined
on the infinite interval $[0,\infty)$ by the Picard-Lindel\"{o}f theorem
\cite{Coddington1972,Khalil2002}. This is summarized in the following lemma.

\begin{lemma}\label{lem:gwp}
The system driven by the gradient of the LJ potential (\ref{eq12}) has a unique
global solution for all $t \geq 0$ and remains in the collision-free open set
$\mathbb{R}^d / \mathcal{C}$ if the initial total energy $E_0 < \infty$.
\end{lemma}

In the next section, Lyapunov convergence to an equilibrium point (modulo
translations) is established under a trajectory boundedness assumption.

\section{LYAPUNOV CONVERGENCE AND EQUILIBRIA}

For the system (\ref{eq12}), a natural Lyapunov function is the total energy
$E$. We have seen that $\dot{E} \leq 0$. Define the set
\begin{align}
\Omega_1
&:= \{(\mathbf{x},\mathbf{v})\in\mathbb{R}^{dN}\times\mathbb{R}^{dN} :
     \dot{E}=0\} \nonumber \\
&= \{(\mathbf{x},\mathbf{v})\in\mathbb{R}^{dN}\times\mathbb{R}^{dN} :
    \mathbf{v}=0\}
\label{eq31}
\end{align}

Note that $\Omega_1$ is not invariant since if the system starts at
$(\mathbf{x},0)$ with $\nabla_{\mathbf{x}} U \neq 0$, then
\begin{equation}
\mathbf{v} = 0, \quad M\dot{\mathbf{v}} = -\nabla_{\mathbf{x}} U \neq 0
\label{eq25}
\end{equation}
where $M$ is the diagonal matrix of masses $m_i$. Therefore, the system
immediately leaves $\Omega_1$. The only trajectories that stay in $\Omega_1$
for all $t \geq 0$ are those that also satisfy $\dot{\mathbf{v}} = 0$, i.e.,
$\mathbf{v} = 0$ and $\nabla_{\mathbf{x}} U = 0$. It follows that the largest
invariant subset of $\Omega_1$ is $\Omega_e$, the set of equilibria.

Note that $E$ is not radially bounded, implying that the set
\begin{equation}
\Omega_0 := \{(\mathbf{x},\mathbf{v})\in\mathbb{R}^{dN}\times\mathbb{R}^{dN} :
             E(\mathbf{x}(t)) \leq E_0\}
\label{eq26}
\end{equation}
is not compact. Thus $\dot{E} \leq 0$ does not, in general, imply that the
trajectories $\mathbf{x}(t)$ of (\ref{eq12}) are bounded, as translations and
cluster drift can prevent compactness. It is suggested in \cite{Rapaport2004}
to use a periodic box or add a small confining potential to ensure radial
boundedness. In this paper, instead of adding a periodic domain
\cite{Rapaport2004} or altering the RHS of (\ref{eq15}), we assume without
loss of generality that all trajectories $\mathbf{x}(t)$ considered are
bounded. Such an assumption is not needed for the velocities since by
(\ref{eq12}) we have
\begin{align}
\frac{1}{2}\sum_{i=1}^N m_i\|\mathbf{v}_i(t)\|^2
&= E(\mathbf{x}(t)) - U(\mathbf{x}(t)) \nonumber \\
&\leq E_0 - \inf_{\mathbf{x}} U(\mathbf{x}(t))
\leq E_0 - a\binom{N}{2},\quad t\geq 0
\label{eq27}
\end{align}
Thus,
\begin{equation}
\|\mathbf{v}_i(t)\|^2 \leq
\frac{2}{m_i}\left[E_0 - a\binom{N}{2}\right],
\quad t\geq 0,\quad i=1,\dots,N
\label{eq28}
\end{equation}

Moreover, since $\dot{E} = -\gamma\sum_{i=1}^N \|\mathbf{v}(t)\|^2 \leq 0$,
the energy $E$ is monotone decreasing and $E \leq E_0$. By the monotone
convergence theorem, there exists a finite limit $E_\infty$ such that
$E(\mathbf{x}(t)) \to E_\infty$. Integrating yields
\begin{equation}
\int_0^\infty \|\mathbf{v}(t)\|^2\,dt = \frac{E_0 - E_\infty}{\gamma} < \infty
\end{equation}
which shows that $\mathbf{v}(t) \to 0$ as $t \to \infty$.

LaSalle's invariance theorem \cite{Khalil2002} states that if a trajectory is
bounded, its positive limit set $L^+$ is nonempty, compact, connected,
invariant, and contained in the largest invariant subset $\Omega_e$ of
$\Omega_1$. That is, $L^+ \subset \Omega_e$, where
\begin{equation}
L^+ = \{(\mathbf{y},0) : \exists\,t_n \to \infty
      \text{ s.t. } \mathbf{x}(t_n) \to \mathbf{y},\;
      \mathbf{v}(t_n) \to 0\}
\label{eq29}
\end{equation}
LaSalle's theorem asserts that every solution $\mathbf{x}(t)$ starting in
$L^+$ approaches $\Omega_e$ as $t \to \infty$, i.e.,
\begin{equation}
d\!\left(\mathbf{x}(t), \Omega_e\right) :=
\inf_{\mathbf{y}\in\Omega_e}\|\mathbf{x}(t)-\mathbf{y}\| \to 0
\quad \text{as} \quad t \to \infty
\label{eq30}
\end{equation}
However, in general LaSalle's theorem does not guarantee that $\mathbf{x}(t)$
will converge to a particular equilibrium point. A counterexample was provided
for a first-order system driven by a gradient of an analytic potential for
$N \geq 2$ in~\cite{Palis1982}.

The situation is different for analytic potentials. In our case $U(\cdot)$ is
analytic on $\mathbb{R}^{dN}/\mathcal{C}$, there exists an equilibrium point
$\mathbf{y}_e \in \Omega_e$ such that $\mathbf{x}(t) \to \mathbf{y}_e$.
Following the proof of Theorem~1.1 in~\cite{HarauxJendoubi1998}, we deduce the
subsequent Theorem:

\begin{theorem}\label{Th1}
Any bounded trajectory $\mathbf{x}(t)$ of the system (\ref{eq12}) on the
collision-free set $\mathbb{R}^{dN}/\mathcal{C}$ converges to an equilibrium
$(\mathbf{x}_e, \mathbf{v}_e) \in \Omega_e$ (modulo translations), satisfying
$\nabla_{\mathbf{x}} U(\mathbf{x}_e) = 0$ and $\mathbf{v}_e = 0$, showing all
agents come to a stop.
\end{theorem}

The proof is omitted due to space limitations but will be provided in a journal
version.

\section{IMPLEMENTATION AND NUMERICAL EXPERIMENTS}

This section validates the theoretical results established in Sections II and
III through systematic numerical experiments. The simulations are designed to
verify: (i) the collision-avoidance guarantee and global well-posedness of
Lemma~\ref{lem:gwp}, (ii) convergence to a single equilibrium modulo
translations as claimed in Theorem~\ref{Th1}, and (iii) the scalability and
practical effectiveness of these properties across agent networks of varying
size.

The numerical integration was performed using the Radau method to solve the
initial value problem with tolerance $10^{-6}$. Unless otherwise stated,
simulations used the parameters detailed in Table~\ref{tab:simulation_parameters}.

\begin{table}[htbp]
\caption{Simulation Parameters}
\label{tab:simulation_parameters}
\centering
\renewcommand{\arraystretch}{1.15}
\begin{tabular}{@{}lc@{}}
\toprule
Parameter & Value \\
\midrule
Agent mass $m$ & 1.0 \\
Potential depth $a$ & 1.0 \\
Damping coefficient $\gamma$ & 0.8 \\
Bubble radius $q$ & 0.25 \\
Collision distance $\sigma$ & 0.50 \\
Equilibrium distance $r^* = 2^{1/6}\sigma$ & 0.5612 \\
\bottomrule
\end{tabular}
\end{table}

\subsection{Two-Agent Convergence: Validation of Global Well-Posedness}

The two-agent case serves as a fundamental validation of Lemma~\ref{lem:gwp},
which guarantees global well-posedness and collision exclusion for
collision-free initial data. For $N=2$, the equilibrium condition (\ref{eq10})
yields a unique equilibrium distance $r^* = 2^{1/6}\sigma = 0.5612$ for
$\sigma = 0.5$.

Agents were initialized at separation $r_{12}(0) = 0.7296 > \sigma$, yielding
initial energy $E_0 = -0.3714$. According to (\ref{eq23}), the theoretical
minimum distance bound is
\[
r_{\min}^{\text{theory}} =
\left(\frac{2a\sigma^{12}}{E_0 + 2a}\right)^{1/12} = 0.5085.
\]

\begin{figure}[t]
\centerline{\includegraphics[width=0.97\columnwidth]{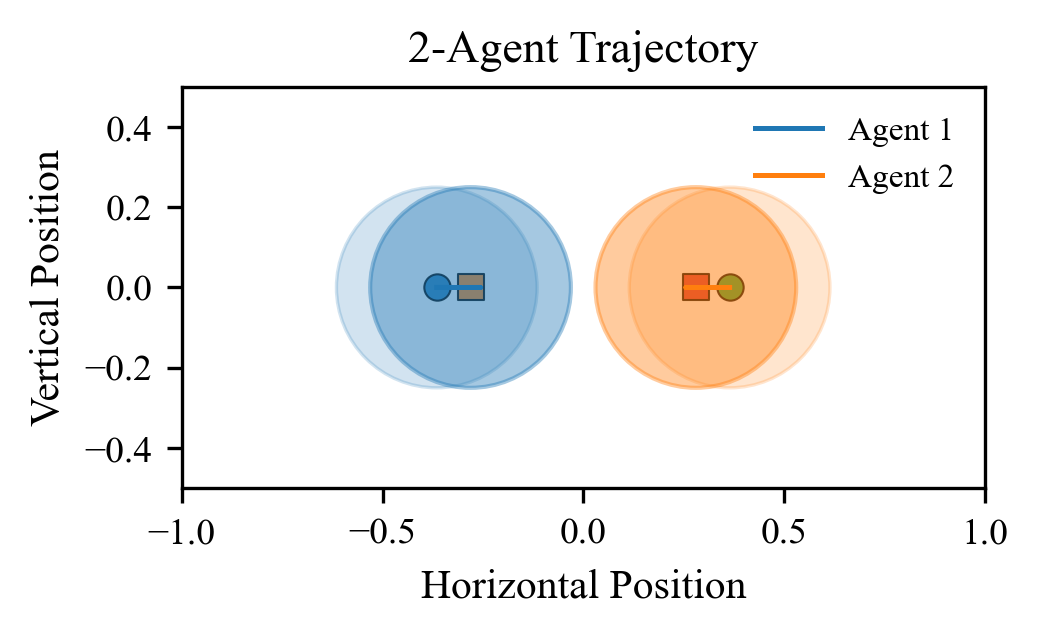}}
\caption{Two-agent trajectories converging to equilibrium distance $r^*$.}
\label{fig1}
\end{figure}

Figure~\ref{fig1} and Figure~\ref{fig4}(b) confirm that $r_{12}(t)$ remains
strictly above both $\sigma$ and $r_{\min}^{\text{theory}}$ throughout the
evolution, validating the collision-avoidance guarantee of Lemma~\ref{lem:gwp}.
The system converges to $r_{12}^{\text{final}} = 0.5614$, yielding a relative
error of $0.023\%$ from the analytical equilibrium, confirming Theorem~\ref{Th1}.

\subsection{Three-Agent Formation Cases}

The three-agent system provides a critical test of Theorem~\ref{Th1}'s claim of
convergence to a single equilibrium modulo translations. Unlike the $N=2$ case,
the $N=3$ system admits multiple equilibrium configurations, including both
rigid (equilateral) and degenerate (collinear) formations, as derived in
Section~II.

We test two scenarios: (i) initial conditions near the equilateral equilibrium,
and (ii) initial conditions near the collinear equilibrium. Theorem~\ref{Th1}
predicts that each trajectory should converge to a single equilibrium
configuration without oscillating between multiple equilibria, despite the
non-convexity of the energy landscape.

Figure~\ref{fig3} illustrates the spatial evolution of the agents. Both
simulations confirm convergence to a single equilibrium configuration. The
equilateral case converged to the symmetric formation with all distances equal
to $r^* = 0.5612$, while the collinear case converged to the asymmetric
equilibrium with $r_{12} = r_{23} = 0.5605$ and $r_{13} = 1.1210$, matching
the analytical prediction within $0.03\%$. Critically, neither trajectory
exhibited oscillation between equilibria, validating the \L{}ojasiewicz-based
argument in Section~III.

\subsection{Validation of Energy Decay and Lyapunov Convergence}

\begin{figure}[t]
\centering
\subfloat[Potential energy evolution]{%
  \includegraphics[width=0.48\columnwidth]{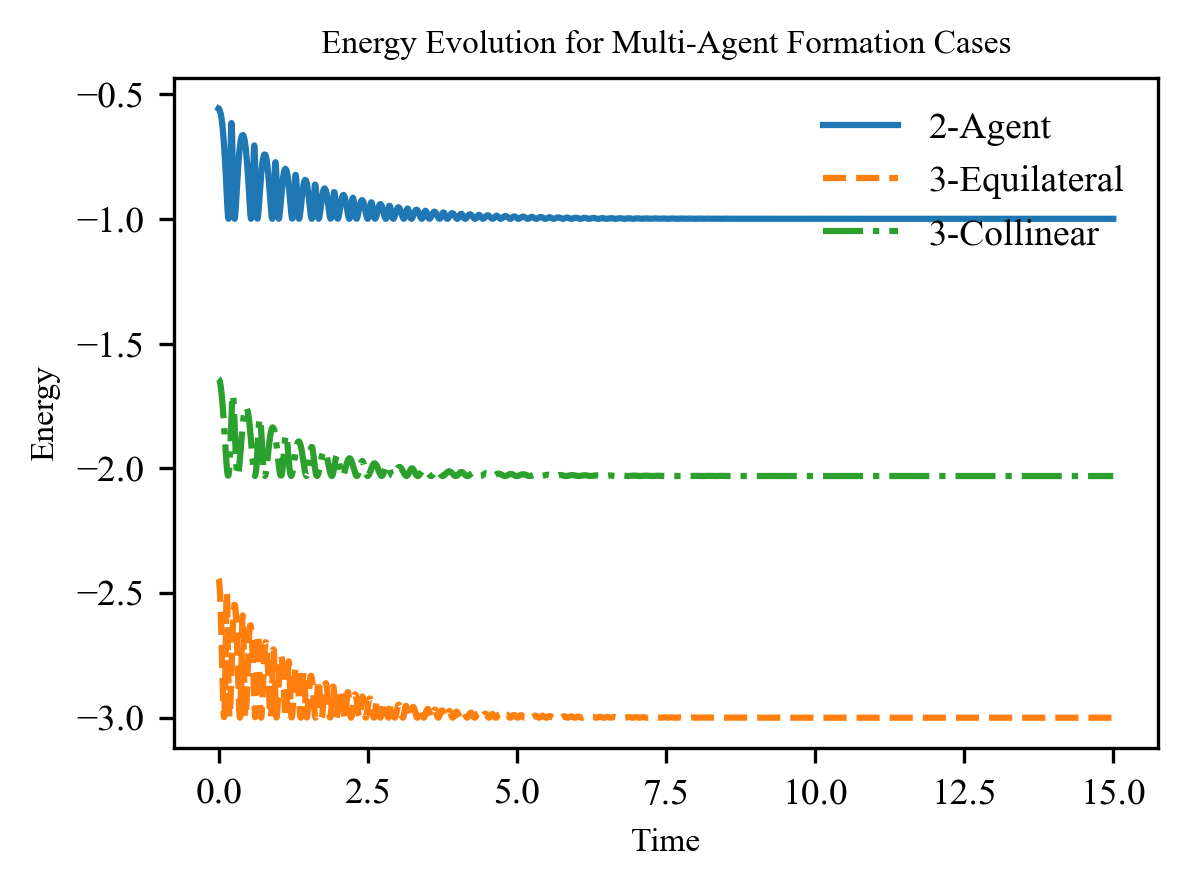}} \hfill
\subfloat[Distance evolution]{%
  \includegraphics[width=0.48\columnwidth]{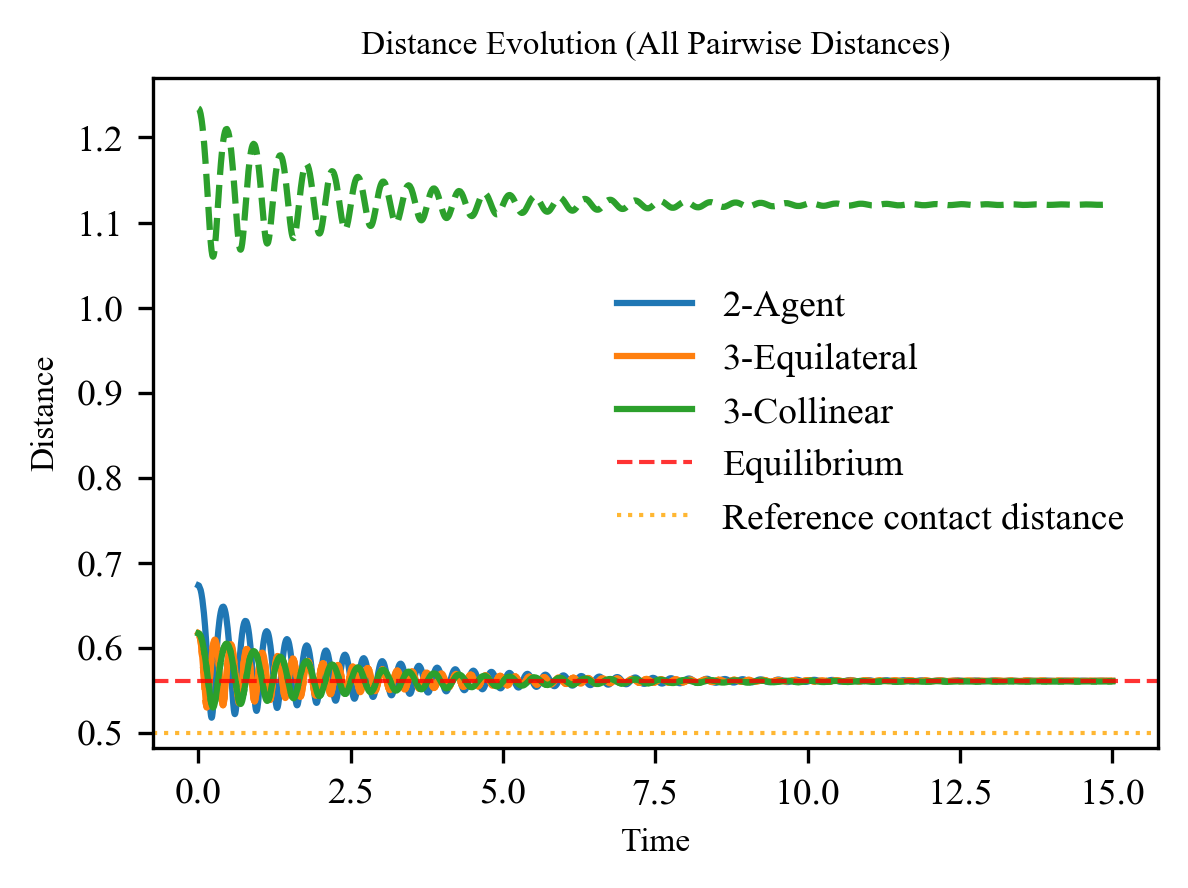}}
\caption{Stability metrics across 2- and 3-agent cases: (a) Potential energy
evolution; (b) All pairwise distances relative to $r^*$ and $\sigma$.}
\label{fig4}
\end{figure}

\begin{figure}[t]
\centering
\subfloat[Equilateral configuration]{%
  \includegraphics[width=0.48\columnwidth]{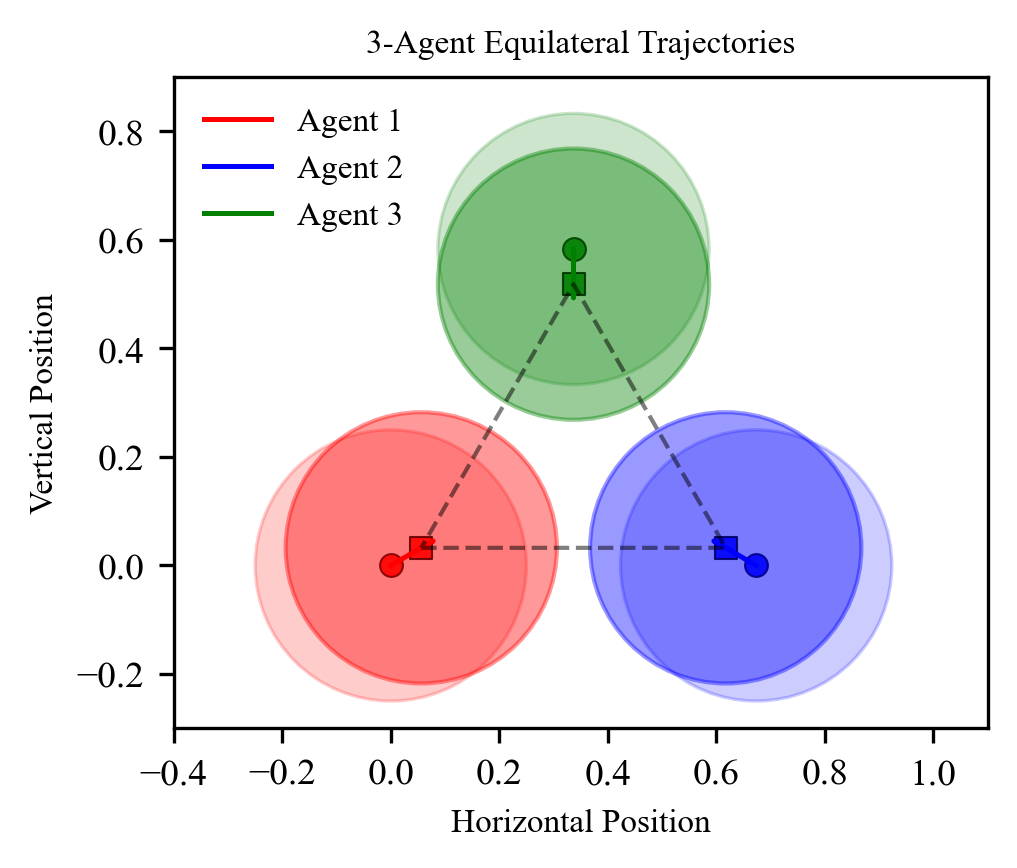}}
\subfloat[Collinear configuration]{%
  \includegraphics[width=0.48\columnwidth]{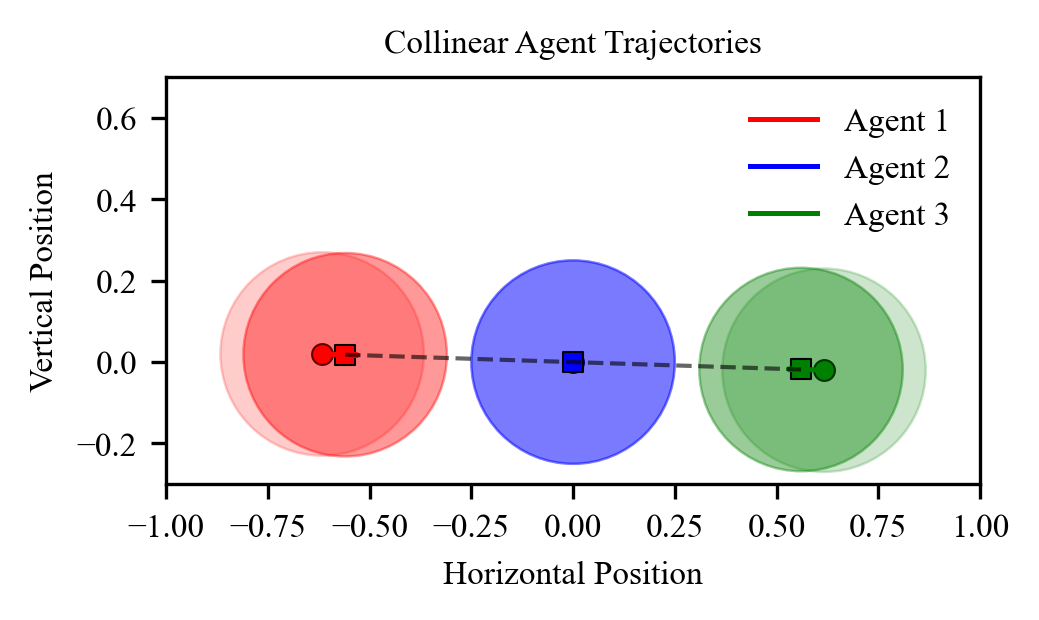}}
\caption{Trajectory plot of 3-agent: (a) Equilateral configuration;
(b) Collinear configuration.}
\label{fig3}
\end{figure}

Figure~\ref{fig4} provides direct validation of the Lyapunov-based convergence
analysis in Section~III. Panel~(a) confirms that the total energy $E(t)$ is
strictly monotone decreasing in all cases, consistent with (\ref{eq28}):
\[
\frac{dE}{dt} = -\gamma\sum_{i=1}^N \|\dot{\mathbf{x}}_i\|^2 \leq 0.
\]
The energy converges to finite limits: $E_\infty = -1.0000$ (2-agent),
$E_\infty = -3.0000$ (equilateral), and $E_\infty = -2.0311$ (collinear).
These values correspond to the potential energy at equilibrium, confirming that
kinetic energy vanishes as $t \to \infty$, as required by Theorem~\ref{Th1}.

Panel~(b) demonstrates that all pairwise distances remain strictly above the
collision threshold $\sigma = 0.5$ throughout the evolution, with minimum
observed distances of $0.5553$ (2-agent), $0.5612$ (equilateral), and $0.5605$
(collinear). In each case, the observed minimum exceeds the theoretical lower
bound (\ref{eq23}), validating Lemma~\ref{lem:gwp}.

\subsection{N-Agent Formation}

To evaluate the scalability of the proposed control strategy, a simulation was
conducted for $N = 8$ agents initialized within the bounded domain
$[-1.0, 1.0]^2$. For $N=8$, the system involves $\binom{8}{2} = 28$ pairwise
interactions. The theoretical minimum distance bound (\ref{eq23}) becomes
\[
r_{\min}^{\text{theory}} =
\left(\frac{2a\sigma^{12}}{E_0 + 2a\cdot 28}\right)^{1/12} = 0.3831.
\]

The observed minimum distance $r_{\min}^{\text{obs}} = 0.5553$ exceeds both
$\sigma = 0.5$ and the theoretical bound, confirming that
Lemma~\ref{lem:gwp}'s collision-avoidance guarantee scales to larger systems.
The distribution of final distances (mean $0.8453$, std $0.2134$) reflects the
emergent geometric structure, with nearest-neighbor distances clustering near
$r^* = 0.5612$ and longer-range distances extending to $1.4745$.

The energy decay from $E_0 = -7.4258$ to $E_\infty = -13.7171$ demonstrates
that the Lyapunov convergence mechanism remains effective for $N=8$, though the
increased dimensionality of the configuration space ($16$-dimensional for $d=2$)
introduces greater geometric complexity in the equilibrium manifold.

\begin{figure}[htbp]
\centerline{\includegraphics[width=0.9\columnwidth]{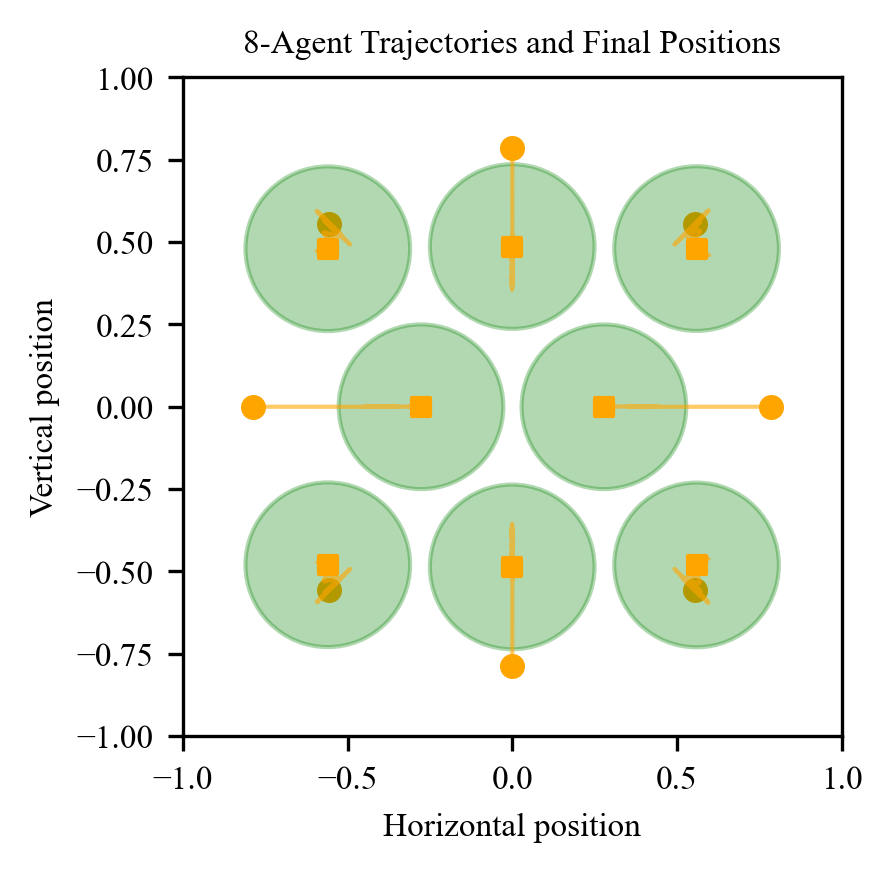}}
\caption{Final configuration and trajectories of $N = 8$ agents.}
\label{fig8}
\end{figure}

\begin{figure}[htbp]
\centering
\subfloat[Total energy evolution]{%
  \includegraphics[width=0.48\columnwidth]{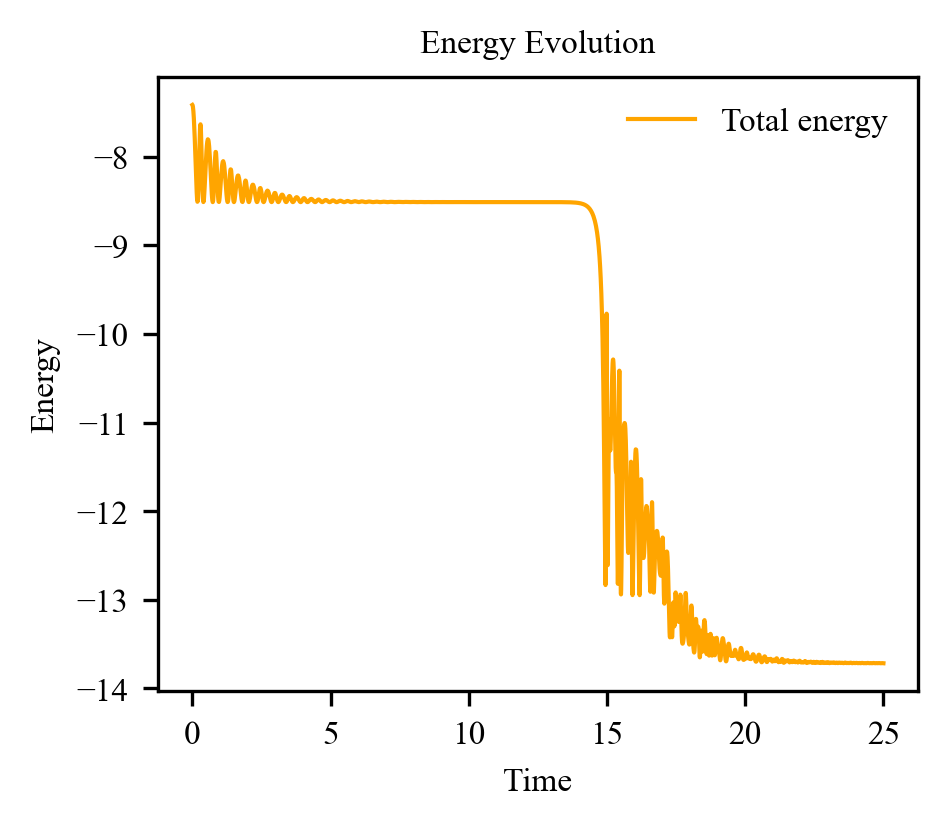}}
\subfloat[Distance evolution]{%
  \includegraphics[width=0.48\columnwidth]{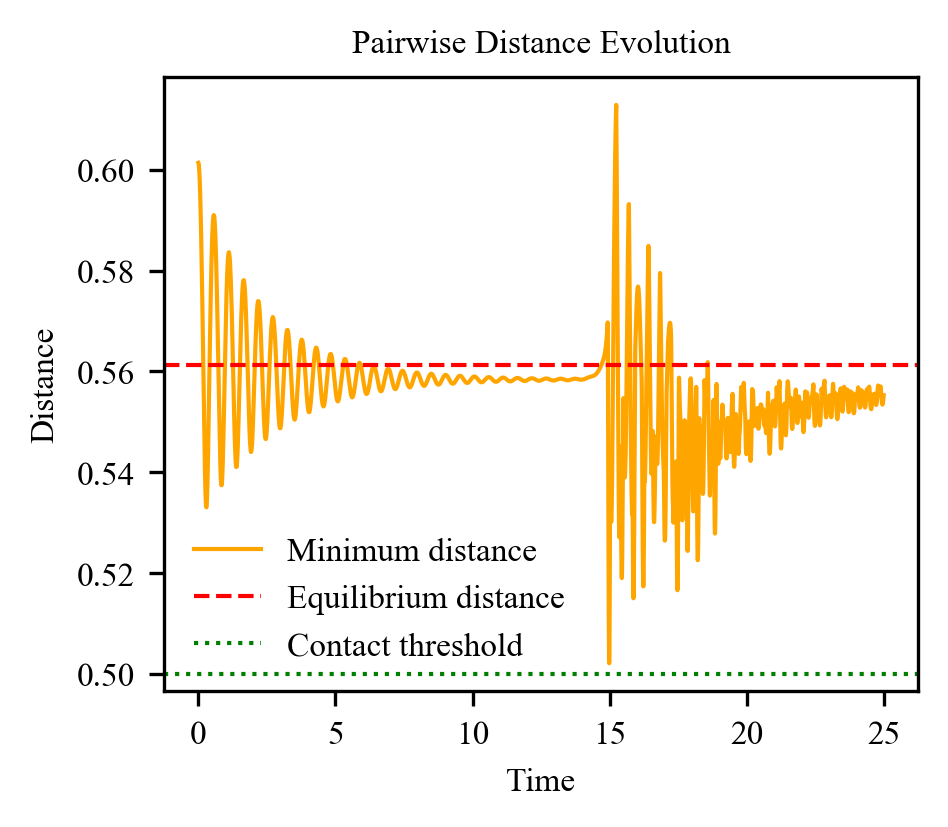}}
\caption{Stability metrics for $N = 8$ agent system: (a) Energy over time;
(b) Pairwise distances.}
\label{fig9}
\end{figure}

\subsection{Validation of Exponential Convergence Rate}

Local linearization of system~\eqref{eq15} around any equilibrium
$(\mathbf{x}^\star, 0)$ yields a damped harmonic oscillator structure, and
standard eigenvalue analysis gives a predicted exponential decay rate
$\alpha = \min\{\gamma/(2m),\, \lambda_{\min}(H)/\gamma\}$,
where $H = \nabla^2_{\mathbf{x}} U(\mathbf{x}^\star)$ \cite{Khalil2002}.
Details of the late-time analysis are omitted here due to page limitations.

For $N=2$ with $m=1.0$ and $\gamma=0.8$, the Hessian at equilibrium
$r^* = 0.5612$ has minimum eigenvalue
$\lambda_{\min}(H) = 2\phi''(r^*) = 457.18$. This yields the predicted rate
\[
\alpha^{\text{theory}}
= \min\!\left\{\frac{0.8}{2\cdot 1.0},\, \frac{457.18}{0.8}\right\}
= \min(0.4,\; 571.47) = 0.4000.
\]

Figure~\ref{fig9} plots $\log(\|r_{12}(t) - r^*\|)$ versus time for
$t \in [0,25]$, after transient effects have decayed. Linear regression yields
an observed decay rate $\alpha^{\text{obs}} = 0.3995$ with $R^2 = 0.9999$,
confirming exponential convergence. The relative error between predicted and
observed rates is $0.125\%$, confirming the local exponential convergence
behavior predicted by the linearized dynamics.

\section{CONCLUSIONS}

This paper establishes a rigorous analytical framework for distributed
multi-agent formation control driven by Lennard--Jones-based
attraction--repulsion potentials. Specifically, we prove global well-posedness
and strict collision avoidance under collision-free initial conditions
(Lemma~\ref{lem:gwp}), establish convergence to a single equilibrium
configuration modulo translations via energy-based Lyapunov analysis and
\L{}ojasiewicz arguments (Theorem~\ref{Th1}), and provide numerical validation
of the theoretical predictions across multiple system scales and formation
topologies.

Section~IV demonstrates that the theoretical guarantees hold in practice.
Simulations confirm: (a)~inter-agent distances remain strictly above the
theoretical lower bound \eqref{eq23} throughout evolution,
(b)~trajectories converge to single equilibria without oscillation between
multiple equilibria, (c)~late-time behavior is consistent with locally
exponential convergence as predicted by the linearized dynamics, and
(d)~collision-avoidance and convergence properties scale effectively to larger
networks. Future work will extend this framework to new applications, such as
power grids, where the distributed attraction-repulsion dynamics have not been
explored.

\addtolength{\textheight}{-12cm}

\bibliographystyle{ieeetr}
\bibliography{ref.bib}

\end{document}